\documentclass[%
prx,aps,twocolumn,secnumarabic, nobibnotes,superscriptaddress
 amsmath,amssymb,
 physrev,
]{revtex4-2}

\usepackage{graphicx}
\usepackage{dcolumn}
\usepackage{bm}
\usepackage{amsfonts,amsmath,amssymb,amsthm}
\usepackage[dvipsnames,usenames]{color}
\usepackage[utf8]{inputenc}
\usepackage[T1]{fontenc}
\usepackage{mathptmx}
\usepackage{etoolbox}
\usepackage{color}
\usepackage{blindtext}
\usepackage{hyperref}
\usepackage{enumerate}
\usepackage[final]{changes}
\hypersetup{colorlinks=true, allcolors=blue}

\begin{document}

\preprint{PRX}

\title{Thermoelectric Optimization and Quantum-to-Classical Crossover in Gate-Controlled Two-dimensional Semiconducting Nanojunctions}

\author{Yu-Chang Chen$^{1,2}$}%
\email{yuchangchen@nycu.edu.tw}
\author{Yu-Chen Chang$^{1}$}
\affiliation{$^{1}$
Department of Electrophysics, National Yang Ming Chiao Tung University, 1001, Daxue Rd., Hsinchu City 300093, Taiwan
}%
\affiliation{$^{2}$
Center for Theoretical and Computational Physics, National Yang Ming Chiao Tung University, 1001, Daxue Rd., Hsinchu City 300093, Taiwan   
}%



\date{\today}

\begin{abstract}
We investigate the thermoelectric performance of Pt–WSe$_2$–Pt nanojunctions with gate-tunable architectures and varying channel lengths from 3 to 12 nm using a combination of first-principles simulations, including density functional theory (VASP), DFT with non-equilibrium Green’s function formalism (NANODCAL), and non-equilibrium molecular dynamics simulations (LAMMPS). Our study reveals a gate- and temperature-controlled quantum-to-classical crossover in electron transport, transitioning from quantum tunneling in short junctions to thermionic emission in longer ones. We observe nontrivial dependencies of the thermoelectric figure of merit (ZT) on the Seebeck coefficient, electrical conductivities, and thermal conductivities as a result of this crossover and gate-controlling. We identify that maximizing ZT requires tuning the chemical potential just outside the band gap, where the system lies at the transition between insulating and conducting regimes. While enormous Seebeck coefficients ($>5000~\mu$V/K) are observed in the insulating state, they do not yield high ZT due to suppressed electrical conductivity and dominant phononic thermal transport. The optimal ZT ($>2.3$) is achieved in the shortest (3 nm) junction at elevated temperatures (500 K), where quantum tunneling and thermionic emission coexist. These findings offer fundamental insights into transport mechanisms in 2D semiconducting nanojunctions and present design principles for high-efficiency nanoscale thermoelectric devices.
\end{abstract}

\maketitle


\section{ Introduction}
\label{sec:Intro}

Thermoelectricity (TE) is a form of sustainable energy capable of converting waste heat into usable electrical power~\cite{Dresselhaus_1}. A central quantity in thermoelectric research is the Seebeck coefficient ($S$), which measures a material’s ability to generate an electrical voltage ($\Delta V$) in response to a temperature difference ($\Delta T$) between two electrodes. It is defined as $S = \frac{\Delta V}{\Delta T}$~\cite{Goldsmid}.

Nanoscale thermoelectric devices represent a novel class of components with the potential for integration into chipsets, enabling enhanced output voltages via series connections of multiple junctions~\cite{Ruitenbeek,YCChen_ACSNano,Amanatidis2015,Majumdar}. Recently, there has been growing interest in the thermoelectric properties of nanojunctions, especially following experimental advances in measuring Seebeck coefficients in atomic and molecular systems~\cite{DubiYonatanandDiVentra}. Recently, we have developed nanojunctions based on nickel-ion-chelated DNA nanowires, in which reversible redox reactions involving electron emission or absorption significantly enhance the Seebeck coefficient. This dynamic Seebeck effect, which is a new type of thermopower, can yield exceptionally large Seebeck coefficients, exceeding $10^5~\mu$V/K~\cite{YCChen_DynamicSeebeck}.

\added{
Two-dimensional transition metal dichalcogenides (2D TMDs) have emerged as a compelling class of materials for thermoelectric applications, combining unique electronic properties with exceptional tunability \cite{Zhang_JMCC_2017,Puthran_JEMRev_2024}. The electronic band structure of 2D TMDs exhibits several advantageous features for thermoelectric applications \cite{Huang_PCCP_2014,Wickramaratne_JCP_2014}. The quantum confinement effect in these atomically thin layers creates a unique density of states (DOS) with sharp features near band edges, which is crucial for achieving high Seebeck coefficients \cite{Hippalgaonkar_PRB_2017}. 
}

Among the various classes of semiconducting field effect transistors (FETs), 2D TMDs have surfaced as viable contenders~\cite{Wang2012, Splendiani, Wu2018_Advanced_electronic_materials}. TMDs function effectively as field-effect transistors  and consist of thin atomically layers with tunable band gaps, which can be modulated via gate voltages~\cite{WBJ_2020,WBJ_2021, Radisavljevic2011, Liu2016_Nano_Lett, Liu2019,YuanLiu}. Despite significant efforts to improve the electronic performance of 2D TMD-based FETs~\cite{Ng_Nature_elecon_2022, Shen_Nature_2021, Wu_Nature_Rev_Mats_2023, Jiang_NatureElectron_2024}, comparatively little research has focused on the thermoelectric properties of nanojunctions with a gate architecture.

\added{
A key advantage of two-dimensional transition metal dichalcogenides (2D TMDs) for thermoelectric applications is their exceptional gate tunability. Owing to their atomically thin structure, these materials are highly responsive to external electric fields, allowing precise modulation of carrier concentration and transport properties \cite{Zhu_EnergyAdv_2023}. This capability enables real-time optimization of thermoelectric performance and opens opportunities for adaptive energy harvesting systems. Importantly, experimental studies have confirmed this tunability: thickness-dependent thermoelectric properties and gate-optimized thermoelectric power factors have been measured for WSe$_2$, demonstrating that electrostatic gating can effectively enhance thermoelectric output \cite{Chen_NanoLett_2023,Yoshida_NanoLett_2016}.
}

In this study, we investigate the thermoelectric efficiency of length-dependent 2D TMD nanojunctions with gate-tunable architectures using first-principles simulations. 
\added{
The present study does not include electron–phonon interactions. However, their influence on the Seebeck coefficient can be significantly enhanced in resonant states \cite{Galperin_JCP_2008}, particularly when the gate voltage shifts the chemical potential outside the band gap region.
}
As the channel length increases, a transition from quantum tunneling to classical thermionic emission is observed.
\added{
When considering gate effects in molecular junctions, a Landauer + DFT approach may encounter fundamental limitations~\cite{Galperin_JCP_2008,Galperin_JPPC_2013}, particularly in short-channel devices. To avoid such issues, we adopt an effective gate model in which the gate-control efficiency is described by $\alpha_{\text{in(out)}}$.
}
In these semiconducting systems, applying a gate voltage ($V_g$) shifts the chemical potential relative to the transmission function $\tau(E)$, thereby inducing a transition from insulating to conducting behavior. The gate architecture thus introduces an additional degree of freedom for tuning electronic transport and optimizing thermoelectric performance.
The ability to control their electronic properties through external gating has opened new avenues for optimizing thermoelectric performance and developing next-generation energy harvesting devices.
The ability to switch between different transport mechanisms and modify carrier concentrations provides multiple pathways for optimizing thermoelectric performance.
\begin{figure*}
\centering
\includegraphics[width=1.0\linewidth]{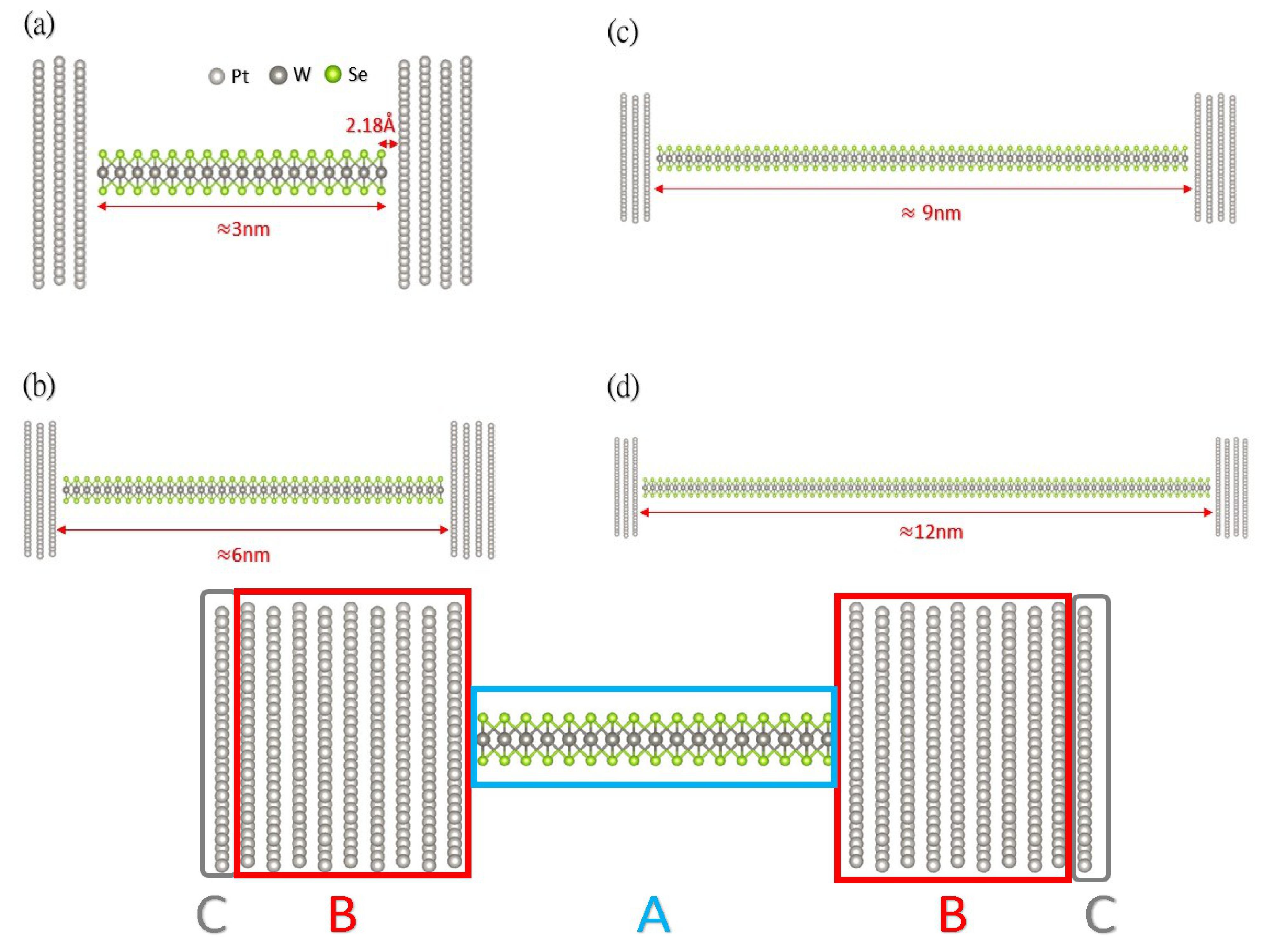}
\caption{\label{fig:Fig1} 
Schematics of Pt-WSe$_{2}$-Pt nanojunctions.
The transmission coefficients $\tau (E)$, conductivity $\sigma$, and electronic thermal conductivity $\kappa_{\mathrm{el}}$ of Pt-WSe$_2$-Pt nanojunctions with channel lengths of (a) 3 nm, (b) 6 nm, (c) 9 nm, and (d) 12 nm are calculated using the NEGF method within the DFT framework with the NANIDcal simulation package. (e) Scheme used to compute the phononic thermal conductivity $\kappa_{\mathrm{ph}}$ via non-equilibrium molecular dynamics (NEMD) using the LAMMPS simulation package. Region A denotes the WSe$_2$ monolayer, regions B represent the thermostatted zones, and atoms in region C are fixed to prevent translational motion.
}
\end{figure*}

We focus on Pt–WSe$_2$–Pt nanojunctions with channel lengths of 3 nm, 6 nm, 9 nm, and 12 nm, as shown in Fig~\ref{fig:Fig1}. Electron transport properties are computed using the NanoDCAL package, which implements the non-equilibrium Green’s function (NEGF) formalism within density functional theory (DFT). For phonon-mediated heat transport, we employ LAMMPS, using non-equilibrium molecular dynamics (NEMD) simulations to calculate the phononic thermal conductivity.

The objective of this work is to elucidate how gate voltage, temperature, and channel length influence the thermoelectric figure of merit, $ZT(T, V_g)$, which arises from a complex interplay among the electrical conductivity $\sigma(T, V_g)$, electronic thermal conductivity $\kappa_{\mathrm{el}}(T, V_g)$, phononic thermal conductivity $\kappa_{\mathrm{ph}}(T)$, and Seebeck coefficient $S(T, V_g)$. We observe that the gate voltage can modulate the nanojunction from p-type to n-type behavior and from an insulating to a conducting state. Additionally, increases in temperature and channel length drive a shift in the dominant electron transport mechanism—from quantum tunneling through the band gap to classical thermionic emission over the barrier.

These combined effects lead to a rich and nontrivial landscape for $ZT(T, V_g)$. Our simulations reveal that Pt–WSe$_2$–Pt nanojunctions can achieve $ZT$ values exceeding 2.6, with the 3 nm junction demonstrating optimal thermoelectric performance at the crossover of the metal–insulator transition and the quantum-to-classical intertwined regime.

\section{Method}
\label{sec:Method}

This study examines Pt-WeSe$_{2}$-Pt junctions with channel lengths of approximately 3 nm, 6 nm, 9 nm, and 12 nm, as illustrated in Fig.~\ref{fig:Fig1}(a)--(d). The junction structures are optimized through energy minimization via density functional theory (DFT) using the VASP simulation package, as detailed in Subsec.~\ref{subsec:VASP}. The transmission coefficient $\tau(E)$ at $V_g=0$ is computed using NEGF-DFT (NANOD), as outlined in Subsection~\ref{subsec:LAMPPS}. The effective gate model $V_{\mathrm{G}}^{\mathrm{eff}}(V_g)$ is detailed in Subsection~\ref{subsec:effective_gate_model}. The phonon's thermal conductivity $\kappa_{\mathrm{ph}}(T)$ is calculated through Non-Equilibrium Molecular Dynamics (NEMD) using LAMMPS, as outlined in Subsection~\ref{subsec:LAMPPS}. 
The theories concerning $\tau$ and $V_{\mathrm{G}}^{\mathrm{eff}}(V_g)$ utilized to compute $\sigma(T,V_g)$, $S(T,V_g)$, $\kappa_{\mathrm{el}}(T,V_g)$, $\kappa_{\mathrm{ph}}(T)$, and $ZT(T,V_g)$ are detailed in Subsec.~\ref{subsec:Thermoelectricity}. 

\subsection{ VASP}\label{subsec:VASP}

The structural optimization and charge density calculations were performed using the Vienna Ab-initio Simulation Package (VASP) \cite{VASP1, VASP2, VASP3}, which solves the Kohn–Sham equations self-consistently within a plane-wave basis set. The Projector-Augmented-Wave (PAW) method \cite{VASP4} was employed as the pseudopotential scheme, and the Perdew–Burke–Ernzerhof (PBE) functional was used to describe exchange-correlation effects \cite{DFT1, DFT2, DFT3, PBE}. We utilized a grid size of $0.016$~\AA$^{-1}$ in reciprocal space and a plane-wave energy cutoff of $400$~eV. The Brillouin zone was sampled using a Monkhorst-Pack $k$-point mesh of [11, 3, 1]. The electronic convergence criterion was set to $10^{-4}$. To ensure high accuracy in the electrostatic potential, particularly in capturing charge transfer effects, a high-density FFT mesh was used. The optimized structures of the Pt electrodes, WSe$_2$ monolayer, and Pt–WSe$_2$–Pt heterojunctions were obtained through structural relaxation using VASP. The [011] crystal plane is rotated by an angle of $30.97^{\circ}$ to minimize the lattice mismatch between the electrode metal region and the WSe$_{2}$ channel region to 0.33\%.

\subsection{ NanoDCAL}\label{subsec:Nanodcal}

Nanodcal (Nanoacademic Device Calculator) is designed for quantitative modeling of quantum transport at the atomic level. The electronic transmission coefficients, $\tau(E)$, are calculated using the NanoDCAL package, which is based on density functional theory (DFT) coupled with the non-equilibrium Green’s function (NEGF) formalism~\cite{Nanodcal1,Nanodcal2,Keldysh}. NanoDCAL performs self-consistent calculations based on the Keldysh nonequilibrium Green's function with the linear combination of atomic orbitals (LCAO) implemented in DFT. Exchange and correlation effects were treated within the local density approximation (LDA). The self-consistent field (SCF) iterations were terminated once the total energy converged to within $10^{-5}$ eV.

\subsection{ LAMPPS}\label{subsec:LAMPPS}

The phonon's thermal conductivity $\kappa_{\mathrm{ph}}(T)$ of the Pt–WSe$_2$–Pt nanojunction is calculated using the Large-scale Atomic/Molecular Massively Parallel Simulator (LAMMPS) package~\cite{PLIMPTON_LAMMPS}. A schematic illustration of the Pt–WSe$_2$–Pt nanojunction is shown in Fig.~\ref{fig:Fig1}(e). The left and right electrodes, each consisting of 570 Pt atoms, act as hot and cold thermal reservoirs, respectively. Interatomic interactions among Pt atoms in the electrode regions are modeled using the Embedded Atom Method (EAM) potential implemented in LAMMPS. The atomic interactions within the WSe$_2$ channel are described by the Stillinger–Weber (SW) potential~\cite{MOBARAKI201892}, while the bonding interactions between the metal electrodes and the WSe$_2$ layer are evaluated via total energy calculations performed with VASP.

We calculated $\kappa_{\mathrm{ph}}(T)$ over a temperature range from 250 K to 500 K. For each temperature, we first performed equilibrium molecular dynamics (EMD) simulations to bring the system to thermal equilibrium. The temperature was controlled using the Berendsen thermostat, with a time step of 0.5 fs, and the simulation was run for $5 \times 10^5$ steps. Once equilibrium was achieved, nonequilibrium molecular dynamics (NEMD) simulations were conducted to compute the heat flux. A temperature gradient was established by controlling the temperatures of the electrodes using the Nose–Hoover thermostat. To generate a thermal current, the temperature of the hot reservoir is set 2.5\% above, and the cold reservoir 2.5\% below, the target temperature at which $\kappa_{\mathrm{ph}}(T)$ is evaluated. Each NEMD simulation is performed with a time step of 0.5 fs for a total of $2 \times 10^7$ steps, with the final $10^6$ steps used to compute the steady-state heat flux.

The thermal conductivity calculated using the NEMD method may exhibit size dependence~\cite{Sellan_2010_PRB}. To assess the effect of system width, we increased the width of the simulation box up to 10 nm and found that the influence on thermal conductivity became negligible beyond this value. Consequently, the width of the Pt–WSe$_2$–Pt simulation box was set to approximately 10 nm in all calculations.

\subsection{ Effective gate model}\label{subsec:effective_gate_model}

First-principles calculations in Ref.~[\cite{FET-AlN}] indicate that the applied gate voltage $V_g$ linearly shifts the chemical potential $\mu$ with respect to $V_g$. It is observed that $V_g$ changes the chemical potential more effectively when the chemical potential is situated within the band gap (i.e., $E_V < \mu < E_C$) than when it is outside the band gap. From the observations from first-principles calculations, we can construct an effective gate model, $V^{\mathrm{eff}}_{\mathrm{G}}(V_g)$, to describe how $\mu$ is shifted by $V_g$ to $\mu(V_g)=\mu+e V^{\mathrm{eff}}_{\mathrm{G}}(V_g) $, as shown in Ref.~[\cite{FET-AlN}]. For $V^{\mathrm{eff}}_{\mathrm{G}}(V_g)$, we remain with the same parameters, $\alpha_{\mathrm{in}}=0.83$ and $\alpha_{\mathrm{out}}=0.33$. The application of $V_g$ causes a shift in $\mu$ by an energy equivalent to 83\% of $e V_g$ when $\mu$ is located within the band gap, while it results in a shift of 33\% of $e V_g$ when $\mu$ is outside the band gap.

According to the effective-gate model and $\tau(E)$ derived from NANODcal, the gate-controllable current is as follows:
\begin{equation} \label{eq:Landauer} 
\begin{aligned}
& I(T_L,T_R,V_{ds},V_g)   = \\
& \frac{2e}{h}\int_{-\infty}^{\infty}{\left[f^R(E,T_R,V_g,V_{ds})-f^L(E,T_L,V_g)\right] \tau \left(E\right) dE},
\end{aligned}      
\end{equation} 
where the Fermi-Dirac distributions of the left and right leads are
\begin{equation} 
f^R(E,T_R,V_g)=\frac{1}{e^{\frac{E-{\mu}_R(V_g,V_{ds})}{k_B T_R}}+1},  \end{equation} 
and,
\begin{equation} 
f^L(E,T_L,V_{ds},V_g)=\frac{1}{e^{\frac{E-{\mu }_L(V_g)}{k_BT_L}}+1}, \end{equation} 
respectively. The chemical potentials of the left and right leads are adjusted by $V_g$ in accordance with the effective gate model, represented as ${\mu }_L\left(V_g\right)=\mu +eV^{\mathrm{eff}}_{\mathrm{G}}(V_g)$ and ${\mu }_R\left(V_g,\ V_{ds}\right)=\mu +e\left[V^{\mathrm{eff}}_{\mathrm{G}}\left(V_g\right)+V_{ds}\right]$, where $V_{ds}$ is the drain-source voltage.

\added{
The correspondence principle states that quantum physics becomes identical to the predictions of classical physics in the limit of high energies and high temperatures. Starting from Landauer formula and $\tau(E) = \frac{h}{L_z} \sum_{n, \mathbf{k}} \delta\left[E - E_{n\mathbf{k}}\right] \frac{|v^z_{n \mathbf{k}}|}{2}$ \cite{EMT-PW}, we have shown that quantum tunneling, as described by the Landauer formula, approaches classical thermionic emission, as given by Richardson’s law, when electrons with energies below the work function are prohibited from tunneling in the classical limit within a free-electron model, as detailed in Subsection D of Section II in Ref.~(\citenum{FET-TMD}).
}

\subsection{ Theory of Thermoelectricity for semiconducting junctions}\label{subsec:Thermoelectricity}

The efficiency of energy conversion in the mono-layered thermoelectric nanojunction, Pt-WSe$_2$-Pt, is characterized by the figure of merit $ZT$, 
\begin{equation} \label{eq:ZT} 
ZT(T,V_g)=\frac{[S(T,V_g)]^2 G(T,Vg)T}{K_{el}(T,V_g)+K_{ph}(T)}, 
\end{equation} 
where the heat conductance transmitted by phonons, denoted as $K_{ph} (T)$, is computed by NEMD utilizing the LAMMPS simulation program. The gate-controllable Seebeck coefficient $S(T,V_g)$, electric conductance $G(T,V_g)$, and thermal conductance $K_{el}(T,V_g)$ attributed to electrons are computed using DFT-NEGF with the Nanodcal package. The details are presented below.

When a small temperature difference $\Delta T=T_R-T_L$ is applied across the electrodes, the Seebeck effect generates a small voltage $\Delta V$ across the electrodes in the gate-controllable thermoelectric nanojunction. Expanding Equation \ref{eq:Landauer} to the lowest order in $\Delta V$ and $\Delta T$ results in: 
\begin{equation} \label{eq:IDTDV}
I(\Delta T,\Delta V,V_g)=G_0 K_0(T,V_g) \Delta V + (\frac{-1}{e T}) G_0
K_1(T,Vg) \Delta T,
\end{equation}
where $G_0=\frac{2e^2}{h}$ is the unit of quantized conductance. Here, we have set $T_L=T$, and $K_n(T,V_g)$ are 
\begin{equation} \label{K_n}
   K_n(T,V_g)=\int_{-\infty}^{\infty} [E-\mu (V_g)]^n \left [-\frac{\partial f(E,T;\mu(V_g))}{\partial E} \right ] \tau(E) dE,
\end{equation}
where $\mu (V_g)=\mu + e V_G^{\mathrm{eff}}(V_g)$. Using Sommerfeld expansion, $K_n(T,V_g)$ can be expressed as a power series in temperature $T$. 
\begin{equation}\label{eq:K_0}
    K_0(T,V_g) \approx \tau[\mu(V_g)]+\frac{\pi^2k_B^2\tau^{[2]}[\mu(V_g)]}{6}T^2
\end{equation} 
\begin{equation}\label{eq:K_1}
    K_1(T,V_g) \approx \frac{\pi^2k_B^2\tau^{[1]}[\mu(V_g)]}{3}T^2,
\end{equation} 
\begin{equation}\label{eq:K_2}
    K_2(T,V_g) \approx \frac{\pi^2k_B^2\tau[\mu(V_g)]}{3}T^2,
\end{equation}
In open circuit, $I(\Delta T,\Delta V,V_g)=0$ and the Seebeck coefficient is define as $S \equiv -\frac{\Delta V}{\Delta T}$. By Eqs.~(\ref{eq:K_0}) and (\ref{eq:K_1}), the Seebeck coefficient is
\begin{equation} \label{eq:S}
\begin{aligned}
     S(T,V_g) &= -\frac{1}{eT} \frac{K_1(T,V_g)}{K_0(T,V_g)} \\ 
    & \approx -\frac{\pi^2 k_B^2}{3e}\frac{\tau^{[1]}[\mu(V_g)]}{\tau[\mu(V_g)]}T,  
\end{aligned}
\end{equation}
where $\tau^{[1]}[\mu(V_g)]$ is the first derivative of $\tau(E)$ at the chemical potential $\mu(V_g)$. 

The electric conductance is defined as $G(T,V_g)\equiv \frac{I(\Delta T,\Delta V,V_g)}{\Delta V} $. As $\Delta V \rightarrow 0$, the electric conductance approaches $G(T,V_g)= G_0 K_0(T,V_g)$, where $G_0=\frac{2e^2}{h}$ is the quantized unit of electrical conductance. When the electric conductance is expanded to the lowest order of temperature, it yields,
\begin{equation}\label{eq:G_T=0}
    G(T,V_g) \approx  \tau[\mu(V_g)]G_0,
\end{equation}
which indicates that the temperature dependence of $G(T,V_g)$ is weak.

The electric conductance $G(T,V_g)$ can be expressed as the sum of contributions from $G_{\mathrm{QM}}(T,V_g)$ and $G_{\mathrm{SC}}(T,V_g)$, such that $G=G_{\mathrm{QM}}+G_{\mathrm{SC}}$, where 
\begin{equation} \label{eq:G_QM}
   G_{\mathrm{QM}}(T,V_g)=G_0\int_{E_V} ^{E_C} \left [-\frac{\partial f(E,T;\mu(V_g))}{\partial E} \right ] \tau(E) dE,
\end{equation}
and 
\begin{equation} \label{eq:G_SC}
   G_{\mathrm{SC}}(T,V_g)=G_0 \left \{ \int_{-\infty} ^{E_V}+\int_{E_C} ^{\infty} \right \} \left [-\frac{\partial f(E,T;\mu(V_g))}{\partial E}\right ] \tau(E) dE,
\end{equation}
where I$G=G_{\mathrm{QM}}$ represents the quantum mechanical conductance component linked to the quantum tunneling current traversing the band gap, a region traditionally deemed forbidden in classical physics.  $G=G_{\mathrm{SC}}$ represents the semiclassical conductance component related to the thermionic emission current.  

From Eqs.~(\ref{eq:G_QM}) and (\ref{eq:G_SC}), we define a parameter $\zeta$ to characterize whether the electron transport mechanism is quantum mechanical or classical:
\begin{equation} \label{eq:zeta}
   \zeta(T,V_g) \equiv\frac{G_{\mathrm{SC}}-G_{\mathrm{QM}}}{G_{\mathrm{SC}}+G_{\mathrm{QM}}},
\end{equation}
where $\zeta>0$ indicates that the primary electron transport mechanism is classical. Conversely, $\zeta<0$ indicates that the primary electron transport mechanism is quantum mechanical.

Likewise, the electron's thermal current that can be controlled by a gate is,
\begin{equation}
\begin{aligned} 
I_Q^{el}(T_L,T_R,V_{ds},V_g)=-\frac{2}{h}\int_{-\infty}^{\infty} & \left[f^R(E,T_R,V_g,V_{ds})-f^L(E,T_L,V_g)\right] \\
& \cdot [E-\mu(V_g)]\tau \left(E\right) dE, 
\end{aligned}
\label{eq:LandauerI_Q}     
\end{equation}

By expanding Eq.~(\ref{eq:LandauerI_Q}) to the lowest order in $\Delta T$, we can determine the electron's thermal conductance as  $K_{el}\equiv\frac{I_Q^{el}}{\Delta T}$,
\begin{equation} \label{K_el}
\begin{aligned}
     K_{el}(T,V_g) &=  \frac{2e}{h}S(T,V_g) K_1(T,V_g) + \frac{2}{hT}K_2(T,V_g)
 \\ 
    & \approx -\left[ \frac{2\pi^2 k_B^2 }{3h}\tau[\mu(V_g)] \right]T ,  
\end{aligned}
\end{equation}

From Eqs.~(\ref{eq:G_T=0}) and (\ref{K_el}), we derive the Wiedemann-Franz law,
\begin{equation}\label{eq:Wiedemann-Franz}
    \frac{K_{el}(T,V_g)}{G(V_g)}=LT,
\end{equation} 
where $L=\frac{\pi^2k_B^2}{3 e^2}$=$2.44\times 10^{-8}$~$\mathrm{W} \Omega \mathrm{K}$ is the Lorentz number

The two-dimensional Pt-WSe$_2$-Pt thermoelectric nanojunctions are periodic in the $x$ direction in our calculations. The definitions of conductivity, electron thermal conductivity, and phonon thermal conductivity are given by $\sigma=\frac{G}{L_x}$, $\kappa_{\mathrm{el}}=\frac{K_{el}}{L_x}$, and $\kappa_{\mathrm{ph}}=\frac{K_{ph}}{L_x}$, respectively, where $L_x$ represents the cross section of 2D WSe$_2$. In this instance, the thermoelectric figure of merit is expressed as $ZT=\frac{S ^2 \sigma}{\kappa_{\mathrm{el}}+\kappa_{\mathrm{ph}}}T$. 

Utilizing the Wiedemann-Franz law, the asymptotic form of $ZT$ is presented as follows:
\begin{equation}
\begin{split}
    ZT(T,V_g) & \approx \frac{S ^2/L}{\left(1+\kappa_{\mathrm{ph}}/\kappa_{\mathrm{el}} \right)}  \\ 
    & \approx \begin{cases}
[S(T,V_g)]^2/L, & \mathrm{if}~ \dfrac{\kappa_{\mathrm{ph}}}{\kappa_{\mathrm{el}}} \ll 1.
\\
\\
    \dfrac{[S(T,V_g)]^2/L}{\left(\frac{\kappa_{\mathrm{ph}}(T)}{\kappa_{\mathrm{el}}(T,V_g)}\right)}, & \mathrm{if}~ \dfrac{\kappa_{\mathrm{ph}}}{\kappa_{\mathrm{el}}} \gg 1.     
    \end{cases}    
\end{split}
\label{eq:ZT_asymp}    
\end{equation} 

\section{Results and Discussion} 
\label{sec:Results}

Using first-principles approaches, we perform density functional theory (DFT) calculations implemented in VASP, nonequilibrium Green’s function formalism within DFT (NEGF-DFT) using NANODCAL, and nonequilibrium molecular dynamics (NEMD) simulations via LAMMPS to investigate 2D Pt–WSe$_2$–Pt nanojunctions with channel lengths of 3, 6, 9, and 12 nm under the influence of gate voltage ($V_g$). The gate voltage, temperature, and channel length serve as key parameters to modulate the energy conversion efficiency, quantified by the thermoelectric figure of merit:
\begin{equation*}
ZT(T,V_g) = \frac{S(T,V_g)^2 \sigma(T,V_g) T}{\kappa_{\mathrm{el}}(T,V_g) + \kappa_{\mathrm{ph}}(T)},     
\end{equation*}
where $S$ is the Seebeck coefficient, $\sigma$ is the electrical conductivity, $\kappa_{\mathrm{el}}$ is the electronic thermal conductivity, and $\kappa_{\mathrm{ph}}$ is the phononic thermal conductivity.

The optimization of $ZT$ depends critically on the interplay among $S$, $\sigma$, $\kappa_{\mathrm{el}}$, and $\kappa_{\mathrm{ph}}$. When the junction is in an insulating state, $\kappa_{\mathrm{ph}} \gg \kappa_{\mathrm{el}}$ and dominates the denominator of $ZT$. In this regime, even a large Seebeck coefficient does not guarantee a high $ZT$ due to strong suppression by phonon thermal conductivity. Conversely, in a conducting state, $\kappa_{\mathrm{el}} \gg \kappa_{\mathrm{ph}}$, and the denominator is dominated by electronic contributions. However, increases in $\sigma$ are often offset by corresponding increases in $\kappa_{\mathrm{el}}$, leading to limited enhancement of $ZT$. Thus, the optimization of $ZT$ is inherently complex due to the competing effects of these four parameters.

In this study, we explore the conditions for optimizing $ZT(T, V_g)$ across a temperature range of 250–500 K and a gate voltage range of –1.5 to 1.5 V for different channel lengths. The gate voltage modulates the chemical potential $\mu(V_g)$ across the band gap, thereby inducing transitions between insulating and conducting states. Additionally, variations in temperature, gate voltage, and channel length drive a crossover in the electron transport mechanism from quantum tunneling to classical thermionic emission. These factors contribute to intricate profiles of $\sigma(T, V_g)$, $S(T, V_g)$, and $\kappa_{\mathrm{el}}(T, V_g)$, making the optimization of $ZT$ a nuanced balance of competing thermoelectric transport properties.

\begin{figure}[h]
\centering
\includegraphics[width=\linewidth]{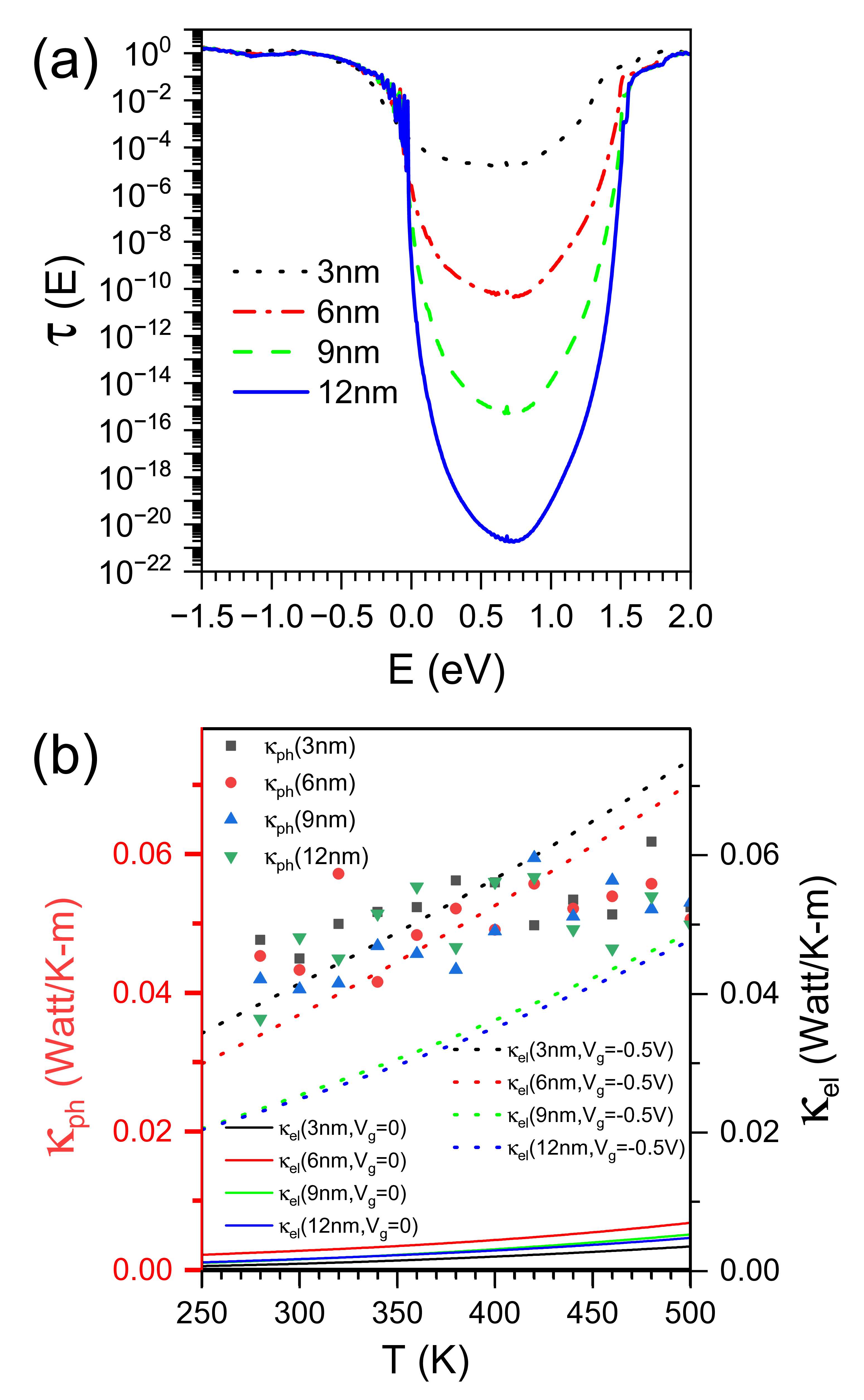}
\caption{
$\tau (E)$ and $\kappa_{ph}(T)$ computed from NANODcal and LAMMPS.
(a) Transmission function $\tau(E)$ as a function of energy $E$ for Pt–WSe$_2$–Pt nanojunctions at $V_g = 0$, with channel lengths $\mathrm{L}_{\text{ch}}=$ 3 nm (black dotted line), 6 nm (red dash-dotted line), 9 nm (green dashed line), and 12 nm (black solid line). The chemical potential $\mu$ at $V_g = 0$ is set to zero and used as the reference energy.
(b) Phononic thermal conductivity $\kappa_{\mathrm{ph}}$ of Pt–WSe$2$–Pt nanojunctions with channel lengths $\mathrm{L}_{\text{ch}}=$ 3 nm (black square), 6 nm (red circle), 9 nm (green upward triangle), and 12 nm (black downward triangle). Solid and dotted lines represent the electronic thermal conductivity $\kappa_{\mathrm{el}}$ at $V_g = 0$ and $V_g = -0.5$ V, respectively, for the same set of channel lengths $\mathrm{L}_{\text{ch}}=$ 3 nm (black), 6 nm (red), 9 nm (green), and 12 nm (blue).
}
\label{fig:Fig2}
\end{figure}

The Pt–WSe$_2$–Pt thermoelectric junction consists of a monolayer of WSe$_2$ serving as the semiconducting channel between two platinum electrodes, as illustrated in Fig~\ref{fig:Fig1}. To analyze electronic transport, we employ the nonequilibrium Green’s function formalism combined with density functional theory (NEGF-DFT) to compute the transmission coefficient $\tau(E)$. Based on $\tau(E)$, the electrical conductivity $\sigma(T, V_g)$, the Seebeck coefficient $S(T, V_g)$, and electronic thermal conductivity $\kappa{\mathrm{el}}(T, V_g)$ are evaluated using the Landauer formalism along with an effective gate model~\cite{FET-AlN}.

As shown in Fig.~\ref{fig:Fig2}(a), the minimum transmission coefficient $\tau_{\mathrm{min}}$ occurs near the center of the band gap. Notably, $\tau_{\mathrm{min}}$ decreases exponentially with increasing nanojunction length, a hallmark of quantum tunneling.

To assess lattice contributions to thermal transport, we perform nonequilibrium molecular dynamics (NEMD) simulations using the LAMMPS package to calculate the phononic thermal conductivity $\kappa_{\mathrm{ph}}(T)$ arising from atomic vibrations. As demonstrated in Fig.~\ref{fig:Fig2}(b), $\kappa_{\mathrm{el}}$ exhibits a linear increase with temperature, consistent with Eq.~(\ref{K_el}), for $T < 350$ K. At higher temperatures ($T > 350$ K), $\kappa_{\mathrm{el}}$ deviates slightly from this linear trend. Compared to $\kappa_{\mathrm{el}}$, the phononic thermal conductivity is less sensitive to temperature changes.

Although $ZT$ scales with $\sigma$, increasing $\sigma$ alone does not guarantee improved thermoelectric performance. This is because $\kappa_{\mathrm{el}}$ typically increases alongside $\sigma$, particularly when $\kappa_{\mathrm{el}} \gg \kappa_{\mathrm{ph}}$ in the metallic regime, thus negating the benefits of enhanced electrical conductivity. Similarly, while $ZT$ is proportional to $S^2$, maximizing $S$ alone does not ensure the highest $ZT$. This behavior can be understood from the approximate expression:
\begin{equation*}
    ZT(T,V_g) \approx \frac{[S(T,V_g)]^2/L}{1+\frac{\kappa_{\mathrm{ph}}(T)}{\kappa_{\mathrm{el}}(T,V_g)}},
\end{equation*}
derived from the Wiedemann–Franz law [Eqs.~(\ref{eq:Wiedemann-Franz}) and (\ref{eq:ZT_asymp})]. The ratio $\kappa_{\mathrm{ph}}/\kappa_{\mathrm{el}}$, which reflects the competition between phononic and electronic heat transport, is a crucial factor governing thermoelectric efficiency. This interplay between the numerator $S^2/L$ and the competing term $1 + \kappa_{\mathrm{ph}}/\kappa_{\mathrm{el}}$ in the denominator determines the optimization of $ZT$.

How this interplay leads to optimization of $ZT(T,V_g)$ to modulated by temperature and gate voltage is illustrated in Fig.~\ref{fig:Fig2}, using the 12 nm Pt–WSe$_2$–Pt nanojunction at $T = 300$ K as a representative example. The left panel of Fig.~\ref{fig:Fig2}(a) exhibits the transmission coefficient $\tau(E)$ as a function of energy $E$ at $V_g = 0$, with the chemical potential set to zero ($\mu = 0$) as the reference energy. The right panel illustrates how the gate voltage $V_g$ shifts the chemical potential from $\mu$ to $\mu(V_g)$, where $\mu(V_g) = \mu + e V{\mathrm{G}}^{\mathrm{eff}}(V_g)$, and $V_{\mathrm{G}}^{\mathrm{eff}}(V_g)$ is determined using the effective gate model \cite{FET-AlN}. Applying a gate voltage of $V_g^{\tau_{\mathrm{min}}}$ shifts the chemical potential to $\mu(V_g^{\tau_{\mathrm{min}}}) = E_{\mathrm{min}}^{\tau}$, where the transmission coefficient $\tau(E)$ reaches its minimum value. This relation is indicated by the horizontal and vertical green dotted lines.

\begin{figure}
\centering
\includegraphics[width=1.0\linewidth]{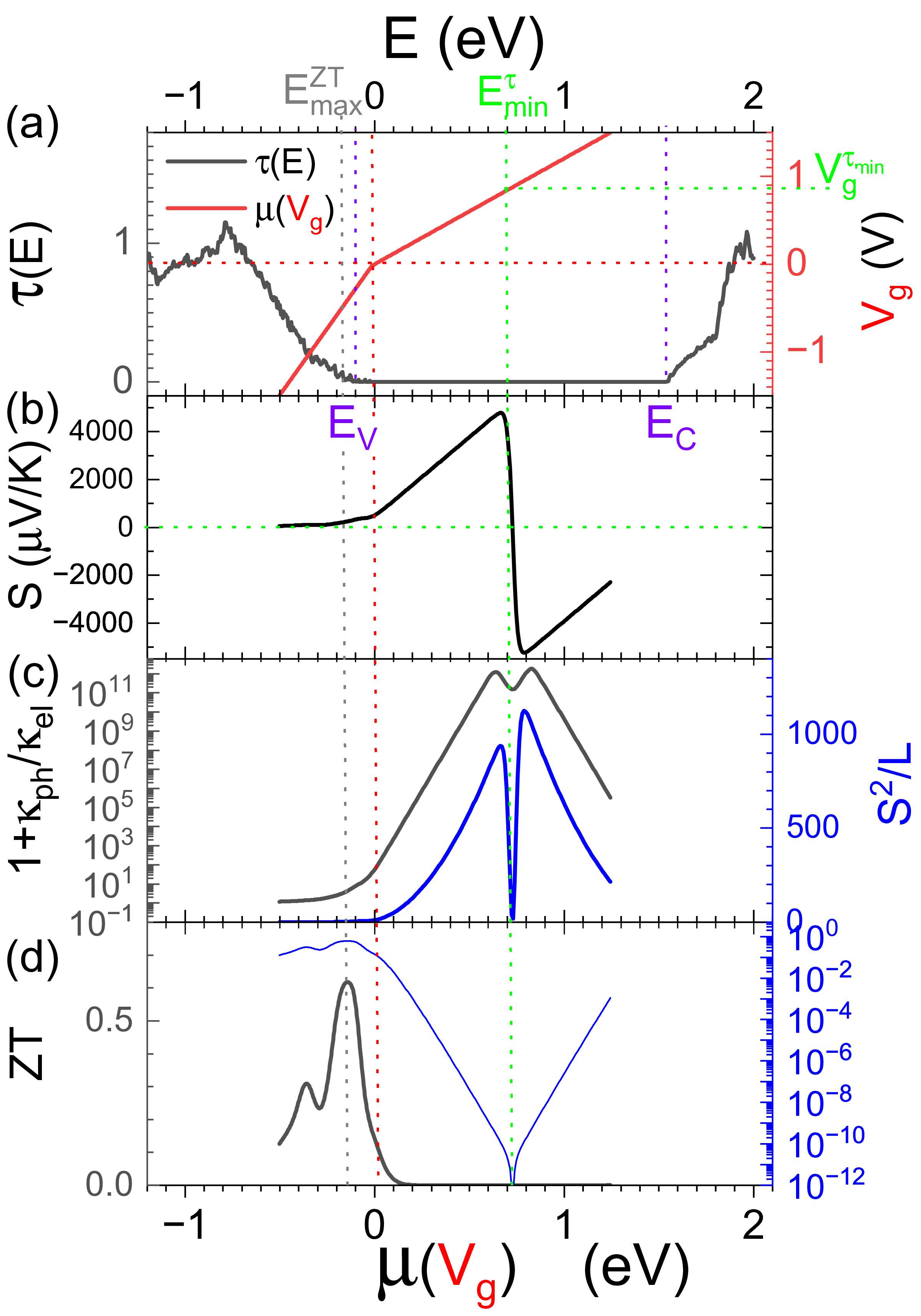}
\caption{
$ZT$ due to competition between $S^2$ and $\kappa_{ph}/\kappa_{el}$.
(a) The black line shows the transmission coefficient $\tau(E)$ as a function of energy $E$ (left vertical axis and top horizontal axis), with the chemical potential $\mu$ set to zero as the reference energy. The red line illustrates how the chemical potential $\mu$ shifts with gate voltage $V_g$ (right vertical axis and bottom horizontal axis), following $\mu(V_g) = \mu + e V_{\mathrm{G}}^{\mathrm{eff}}(V_g)$.
(b) The Seebeck coefficient $S$ is plotted as a function of $\mu(V_g)$, showing its evolution with varying $V_g$.
(c) The factor $1 + \kappa_{\mathrm{ph}}/\kappa_{\mathrm{el}}$ (left axis) and the factor $S^2/L$ (right axis) are plotted as functions of $\mu(V_g)$, highlighting the competition between phononic and electronic thermal transport and the trade-off between maximizing $S$ and minimizing the denominator in $ZT$.
(d) The resulting thermoelectric figure of merit $ZT$ [left (right) axis: linear (log) scale] is plotted as a function of $\mu(V_g)$, reflecting the net outcome of the interplay among $S$, $\kappa_{\mathrm{el}}$, and $\kappa_{\mathrm{ph}}$.
The chemical potential $\mu$ at $V_g=0$ is designated as the reference energy, set to zero, while the transmission band gap is defined by $(E_{\mathrm{V}},E_{\mathrm{C}})$.
}
\label{fig:Fig3}
\end{figure}
Figure~\ref{fig:Fig3}(b) shows the Seebeck coefficient $S$ as a function of the chemical potential $\mu(V_g)$ for the 12 nm Pt–WSe$_2$–Pt junction at $T = 300$ K. Since $S \propto -\tau'(E)/\tau(E)$ [c.f. Eq.(\ref{eq:S})], indicating that the Seebeck coefficient vanishes at $V_g = V_g^{\tau{\mathrm{min}}}$, where the first derivative of the transmission function satisfies $\tau'(E) = 0$.  

When $V_g$ experiences a slight deviation from $V_g^{\tau{\mathrm{min}}}$, the Seebeck coefficients exhibit a sharp increase, surpassing 5000~$\mu$V/K. This behavior can be ascribed to the circumstances under which $V_g$ shifts the chemical potential $\mu(V_g)$, positioning it at an energy close to the local minimum of $\tau(E)$, where $\tau'(E) \neq 0$ and $\tau(E)<10^{-20}$ is notably diminutive. As indicated in Fig.~\ref{fig:Fig3}(d), the substantial magnitude of $S$ corresponds to a minimal $ZT$. This observation suggests that while the nanojunction can produce a significant voltage through the temperature gradient, the efficiency of the thermoelectric battery remains low due to the high internal resistance when the chemical potential is situated at the midpoint of the band gap. In addition, a P-type thermoelectric junction ($S > 0$) is formed when the gate voltage is less than or equal to $V_g^{\tau_{\mathrm{min}}}$, where the transmission function exhibits a positive slope, i.e., $\tau'(E_{\mathrm{min}}^{\tau}) > 0$. As the gate voltage increases beyond $V_g^{\tau_{\mathrm{min}}}$, the slope becomes negative, corresponding to N-type behavior ($S < 0$).

The diminutive $ZT$ linked to the substantial magnitude of $S$ in the bangap area may also be elucidated by $ZT \approx \frac{S^2/L}{1 + \kappa_{\mathrm{ph}}/\kappa_{\mathrm{el}}}$ [cf. Eq.~(\ref{eq:ZT_asymp})], where $L$ denotes the Lorentz number. Figure~\ref{fig:Fig3}(c) displays the denominator $1 + \kappa_{\mathrm{ph}}/\kappa_{\mathrm{el}}$ on the left vertical axis and the numerator $S^2/L$ on the right vertical axis. When $S^2/L$ reaches its extrema in the insulating regime, the denominator also becomes extremely large due to the fact that $\kappa_{\mathrm{ph}} \gg \kappa_{\mathrm{el}}$, reflecting the low electronic thermal conductivity in this regime. As a result, $ZT$ remains very small when the junction is tuned to an insulating state, regardless of how large the Seebeck coefficient becomes.

Figure~\ref{fig:Fig3}(d) shows that $ZT$ reaches its maximum value of 0.617 when the gate voltage $V_g$ shifts the chemical potential to $\mu(V_g) = E_{\mathrm{min}}^{ZT}$, as indicated by the vertical grey dotted line. The corresponding Seebeck coefficient at this point is only $S \approx 221.7~\mu$V/K, which is substantially lower than the peak value of $S \approx 4786~\mu$V/K observed in the P-type regime. The associated ratio $\kappa_{\mathrm{ph}}/\kappa_{\mathrm{el}} \approx 3.61$ indicates that the phononic and electronic contributions to thermal conductivity are comparable, placing the system in the crossover regime between insulating and conducting states.

We observe that $ZT$ reaches its maximum value of 0.617 when the gate voltage $V_g$ shifts the chemical potential to $\mu(V_g) = E_{\mathrm{min}}^{ZT}$, as indicated by the vertical grey dotted line. Notably, the corresponding Seebeck coefficient at this point is only $S \approx 221.7~\mu$V/K, which is significantly lower than the peak value of $S \approx 4786~\mu$V/K observed in the P-type regime. The corresponding ratio $\kappa_{\mathrm{ph}}/\kappa_{\mathrm{el}} \approx 3.61$ suggests that the phononic and electronic contributions to thermal conductance are comparable, indicating that the system is in the transitional regime between insulating and conducting states.

From the perspective of the transmission coefficient $\tau(E)$, the optimal condition for maximizing $ZT$ occurs when the chemical potential lies slightly outside the band gap, at $E = E_{\mathrm{min}}^{ZT}$. At this energy, $\tau(E)$ begins to rise sharply, signaling a transition from an insulating to a conducting state. It is within this transitional regime that the balance among the Seebeck coefficient $S$, electronic thermal conductivity $\kappa_{\mathrm{el}}$, and phonon thermal conductivity $\kappa_{\mathrm{ph}}$ creates the most favorable conditions for enhancing the thermoelectric figure of merit $ZT$.

\begin{figure*}
\centering
\includegraphics[width=1.0\linewidth]{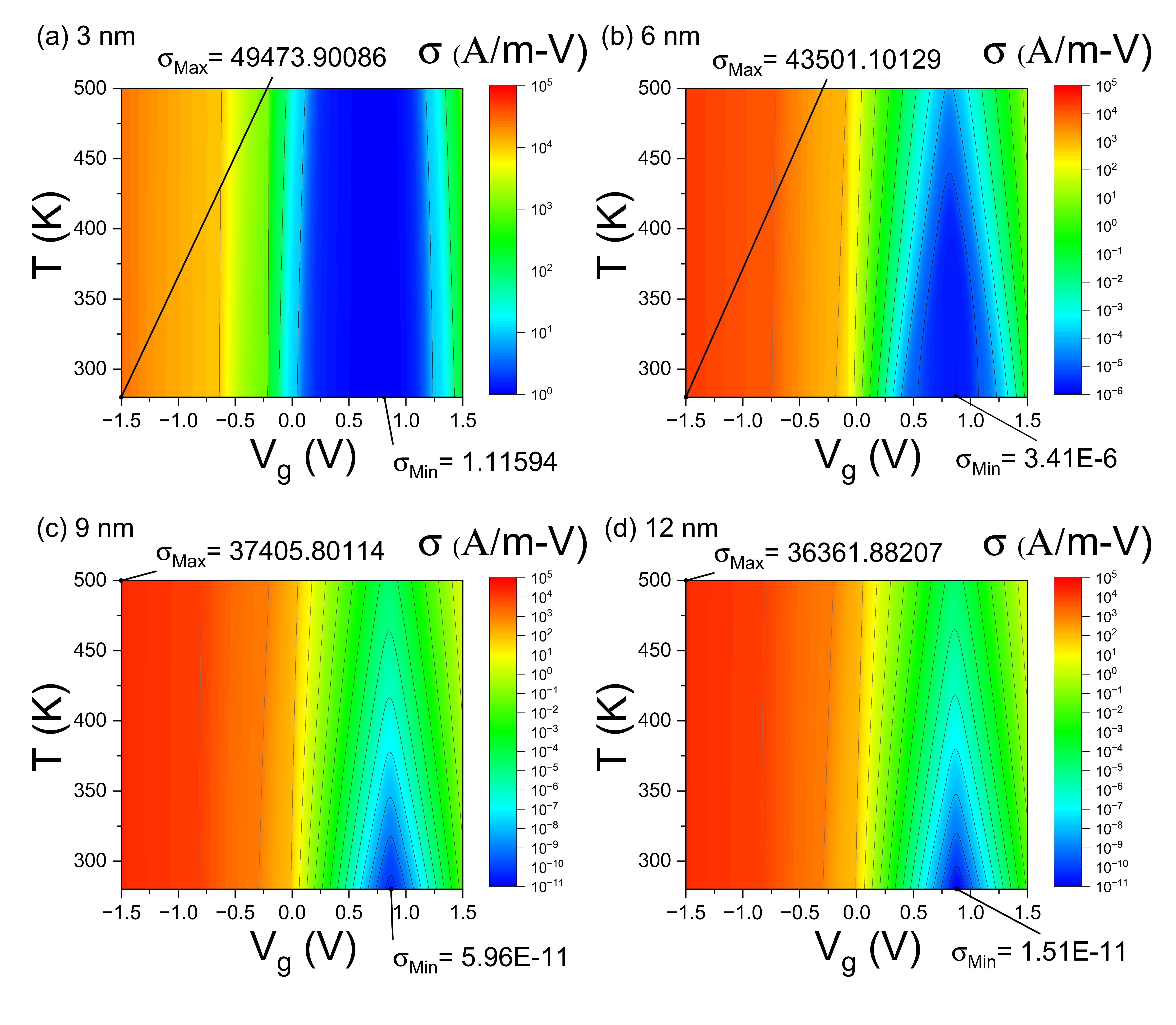}
\caption{ 
Contour plots of $\sigma(T,V_g)$.
Contour plots of the electrical conductivity $\sigma(T, V_g)$ are shown as functions of temperature (250–500 K) and gate voltage $V_g$ (–1.5 to 1.5 V) for Pt–WSe$_2$–Pt thermoelectric junctions with channel lengths $\mathrm{L}_{\text{ch}}=$ (a) 3 nm, (b) 6 nm, (c) 9 nm, and (d) 12 nm. The maximum and minimum values of $\sigma$, denoted as $\sigma_{\mathrm{Max}}$ and $\sigma_{\mathrm{Min}}$, respectively, and their corresponding locations within the $T–V_g$ domain are also indicated.
}
\label{fig:Fig4}
\end{figure*}

Figure~\ref{fig:Fig4} presents contour plots of the electrical conductivity $\sigma(T, V_g)$ for Pt–WSe$_2$–Pt thermoelectric junctions with varying channel lengths over a temperature range of 250-500 K and gate voltage range of $-1.5$ to $1.5$ V. When the gate voltage $V_g$ shifts the chemical potential $\mu(V_g)$ outside the band gap, the junction enters a conducting metallic state, with the corresponding maxima of $\sigma(T, V_g)$ denoted as $\sigma_{\mathrm{Max}}$. Conversely, when $V_g$ shifts $\mu$ to $\mu(V_g)$, which lies within the band gap, the junction becomes insulating, and the corresponding minima are denoted as $\sigma_{\mathrm{Min}}$. As illustrated in Fig.~\ref{fig:Fig4}, $\sigma_{\mathrm{Min}}$ exhibits an exponential decrease with increasing $L_{\mathrm{ch}}$ for nanojunctions where $L_{\mathrm{ch}} < 9$ nm, suggesting that quantum tunneling is the predominant mechanism of electron transport in this particular regime.

Equation~(\ref{eq:G_T=0}) shows the conductance $G=G_0 \tau [\mu(V_g)]$ at $T = 0$ K. Consequently, the lowest value of electric conductivity $\sigma_{\mathrm{Min}}$ at $T=0$ K ought to be proportional to the minimum of the transmission coefficient $\tau_{\mathrm{Min}}$. Due to quantum tunneling, as illustrated in Fig.~\ref{fig:Fig2}(a), the value of $\tau_{\mathrm{Min}}$ drops exponentially as the channel length $\mathrm{L}_{\text{ch}}$ increases.  Consequently, $\sigma_{\mathrm{Min}}$ at $T=0$ K should decrease exponentially as the channel length $\mathrm{L}_{\text{ch}}$ increases. Nonetheless, $\sigma_{\mathrm{Min}}$ at $T=250$ K does not exhibit an exponential decline when the channel lengths range from 9 nm [$\sigma_{\mathrm{Min}}= 5.96 \times 10^{-11}$ A/(m-V)] to 12 nm [$\sigma_{\mathrm{Min}}=1.51 \times 10^{-11}$ A/(m-V)]. This suggests that the junction undergoes a change in the electron transport mechanism from quantum tunneling to classical thermionic emission for nanojunctions where $\mathrm{L}_{\text{ch}}>9$ nm. 

\begin{figure*}
\centering
\includegraphics[width=1.0\linewidth]{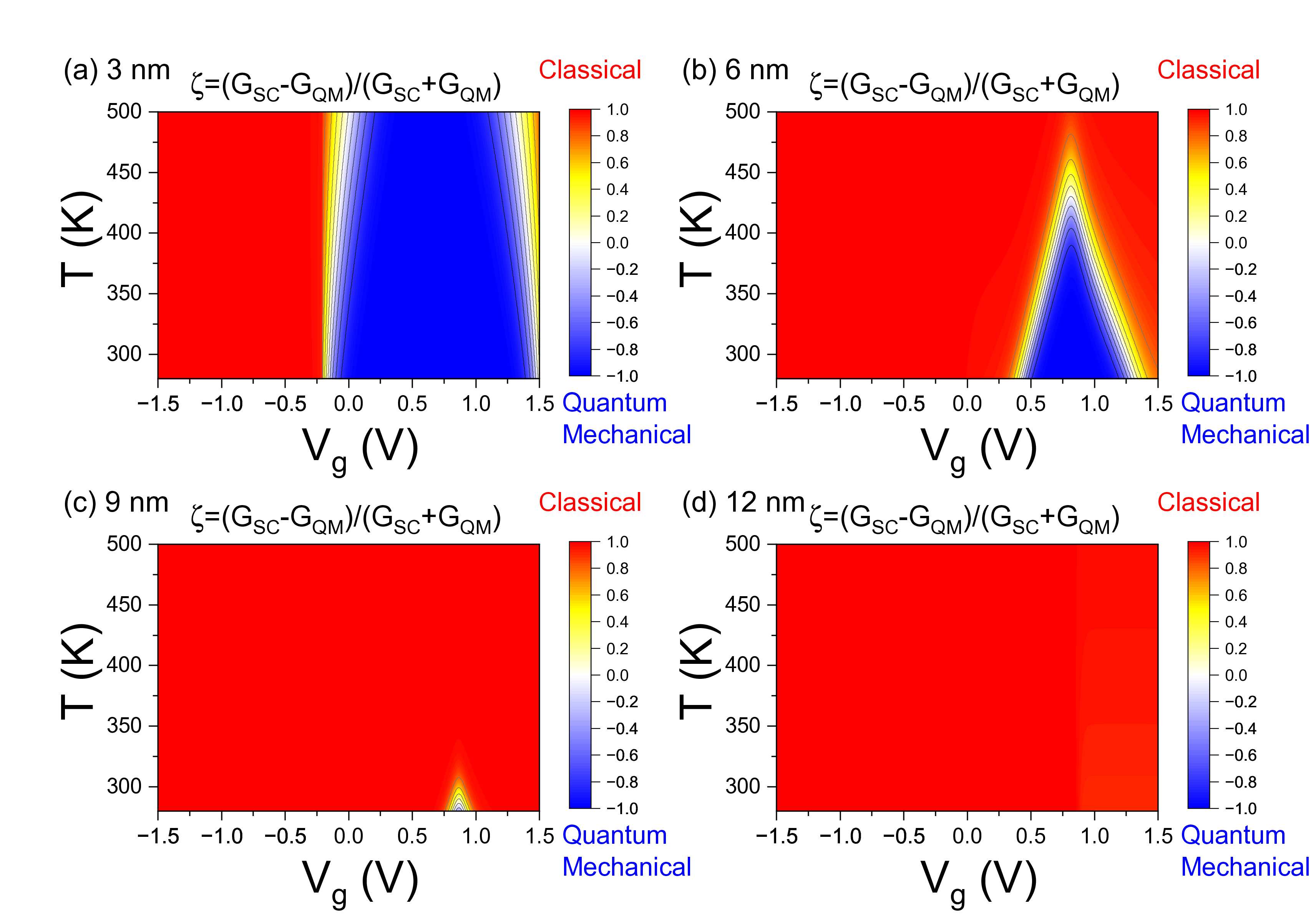}
\caption{ 
Contour plots depict the quantum-to-classical transition.
Contour plots of $\zeta \equiv (G_{\mathrm{SC}}-G_{\mathrm{QM}})/(G_{\mathrm{SC}}+G_{\mathrm{QM}})$ are shown as functions of temperature (250–500 K) and gate voltage $V_g$ (–1.5 to 1.5 V) for Pt–WSe$_2$–Pt thermoelectric junctions with channel lengths $\mathrm{L}_{\text{ch}}=$ (a) 3 nm, (b) 6 nm, (c) 9 nm, and (d) 12 nm. $\zeta(T,V_g)$ represents competitive strength between quantum mechanical (shown in blue) and semi-classical (represented in red) transport mechanisms.
}
\label{fig:Fig5}
\end{figure*}

To illustrate the transition in the electron transport mechanism from quantum tunneling to classical thermionic emission, we define a parameter $\zeta \equiv \frac{G_{\mathrm{SC}}-G_{\mathrm{QM}}}{G_{\mathrm{SC}}+G_{\mathrm{QM}}}$ in Eq.~(\ref{eq:zeta}) to characterize the competitive strength between quantum mechanical and semi-classical transport mechanisms. Figure \ref{fig:Fig5} illustrates that quantum mechanical transport regimes (indicated in blue) diminish from 3 nm to 9 nm and are entirely absent at 12 nm.  

\begin{figure*}
\centering
\includegraphics[width=1.0\linewidth]{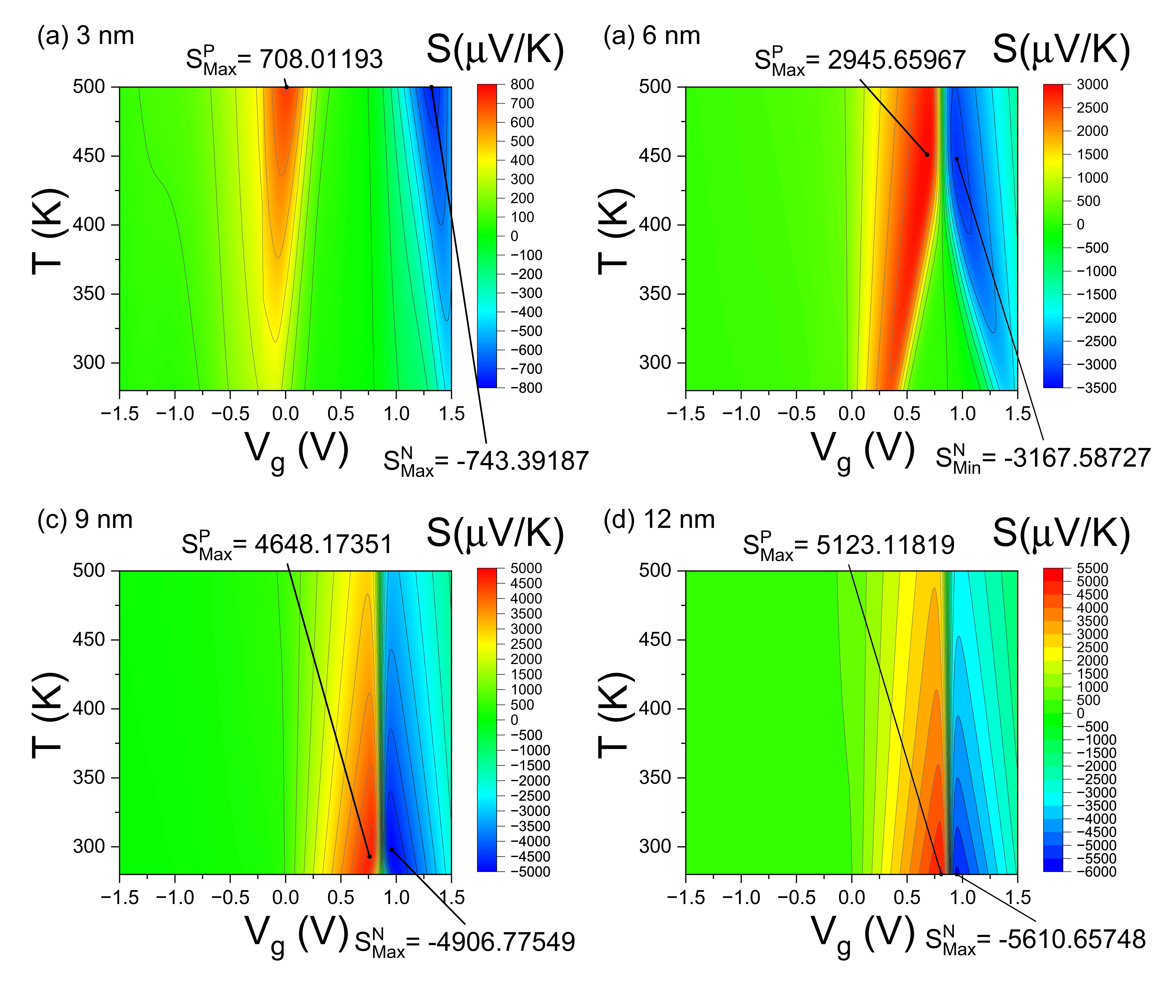}
\caption{ 
Contour plots of $S(T,V_g)$.
Contour plots of the Seebeck coefficient $S(T, V_g)$ are shown over a temperature range of 250–500 K and gate voltage range of –1.5 to 1.5 V for Pt–WSe$_2$–Pt thermoelectric junctions with channel lengths $\mathrm{L}_{\text{ch}}=$ (a) 3 nm, (b) 6 nm, (c) 9 nm, and (d) 12 nm. The maximum absolute values of the Seebeck coefficient in the P-type and N-type regimes are denoted as $S{\mathrm{Max}}^{\mathrm{P}}$ and $S_{\mathrm{Max}}^{\mathrm{N}}$, respectively, along with their corresponding locations within the $T$–$V_g$ domain.
}
\label{fig:Fig6}
\end{figure*}

Figure~\ref{fig:Fig6} presents contour plots of the Seebeck coefficient $S(T, V_g)$ for Pt–WSe$_2$–Pt thermoelectric junctions with various channel lengths. Since $S \propto -\tau'(E)/\tau(E)$, the Seebeck coefficient vanishes at $V_g = V_g^{\tau_{\mathrm{min}}}$, where the transmission function $\tau[\mu(V_g^{\tau_{\mathrm{min}}})]$ reaches a local minimum and its derivative $\tau'[\mu(V_g^{\tau_{\mathrm{min}}})]$ vanishes. For gate voltages below this point ($V_g < V_g^{\tau_{\mathrm{min}}}$), the junction is P-type ($S > 0$); for $V_g > V_g^{\tau_{\mathrm{min}}}$, $S < 0$, indicating N-type behavior.

The extrema of the Seebeck coefficient, denoted as $S_{\mathrm{Max}}^{\mathrm{P}}$ and $S_{\mathrm{Max}}^{\mathrm{N}}$, occur in the vicinity of $V_g^{\tau_{\mathrm{min}}}$ where the transmission $\tau(E)$ is small. These extrema appear when the chemical potential is tuned to lie near the middle of the band gap, placing the junctions in an insulating state. We observe that the magnitudes of both $S_{\mathrm{Max}}^{\mathrm{P}}$ and $S_{\mathrm{Max}}^{\mathrm{N}}$ increase with channel length.

Moreover, distinct differences can be seen between the contour plots of $S(T, V_g)$ for junctions dominated by quantum tunneling and those dominated by classical thermionic emission. Similar to the trends observed in the electrical conductivity $\sigma(T, V_g)$, the $S(V_g)$ profile for the 12 nm junction closely resembles that of the 9 nm junction, further indicating a transition in the dominant transport mechanism around this channel length.

\begin{figure*}
\centering
\includegraphics[width=1.0\linewidth]{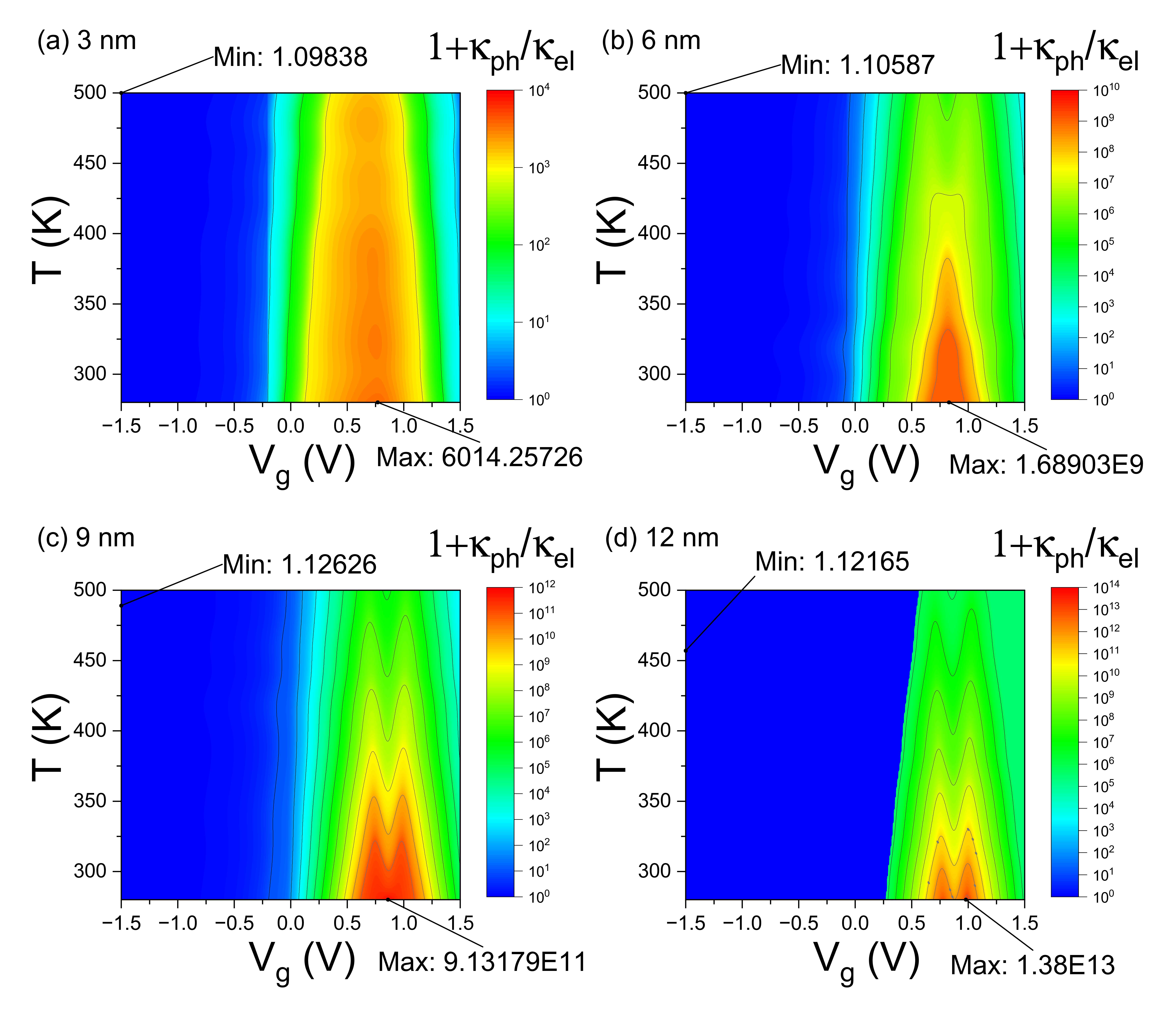}
\caption{ 
Contour plots representing competition between phononic and electronic thermal conductivity.
Contour plots of the factor $1 + \frac{\kappa_{\mathrm{ph}}(T)}{\kappa_{\mathrm{el}}(T, V_g)}$ are shown over the temperature range of 250–500 K and gate voltage range of –1.5 to 1.5 V for Pt–WSe$_2$–Pt thermoelectric junctions with channel lengths of (a) 3 nm, (b) 6 nm, (c) 9 nm, and (d) 12 nm. The factor's maximum and minimum values, along with their respective locations in the $T$–$V_g$ domain, are labeled as Max and Min, respectively.
}
\label{fig:Fig7}
\end{figure*}

Since the thermoelectric figure of merit is approximated by $ZT \approx \frac{S^2/L}{1 + \kappa_{\mathrm{ph}}/\kappa_{\mathrm{el}}}$, the efficiency of energy conversion is governed by a subtle competition between the numerator $S^2/L$ and the denominator $1 + \kappa_{\mathrm{ph}}/\kappa_{\mathrm{el}}$. A large Seebeck coefficient $S$ alone does not necessarily result in a high $ZT$ value. To illustrate this point, we present contour plots of the factor $1 + \kappa_{\mathrm{ph}}/\kappa_{\mathrm{el}}$ in Fig.~\ref{fig:Fig7}. 

\begin{figure*}
centering
\includegraphics[width=1.0\linewidth]{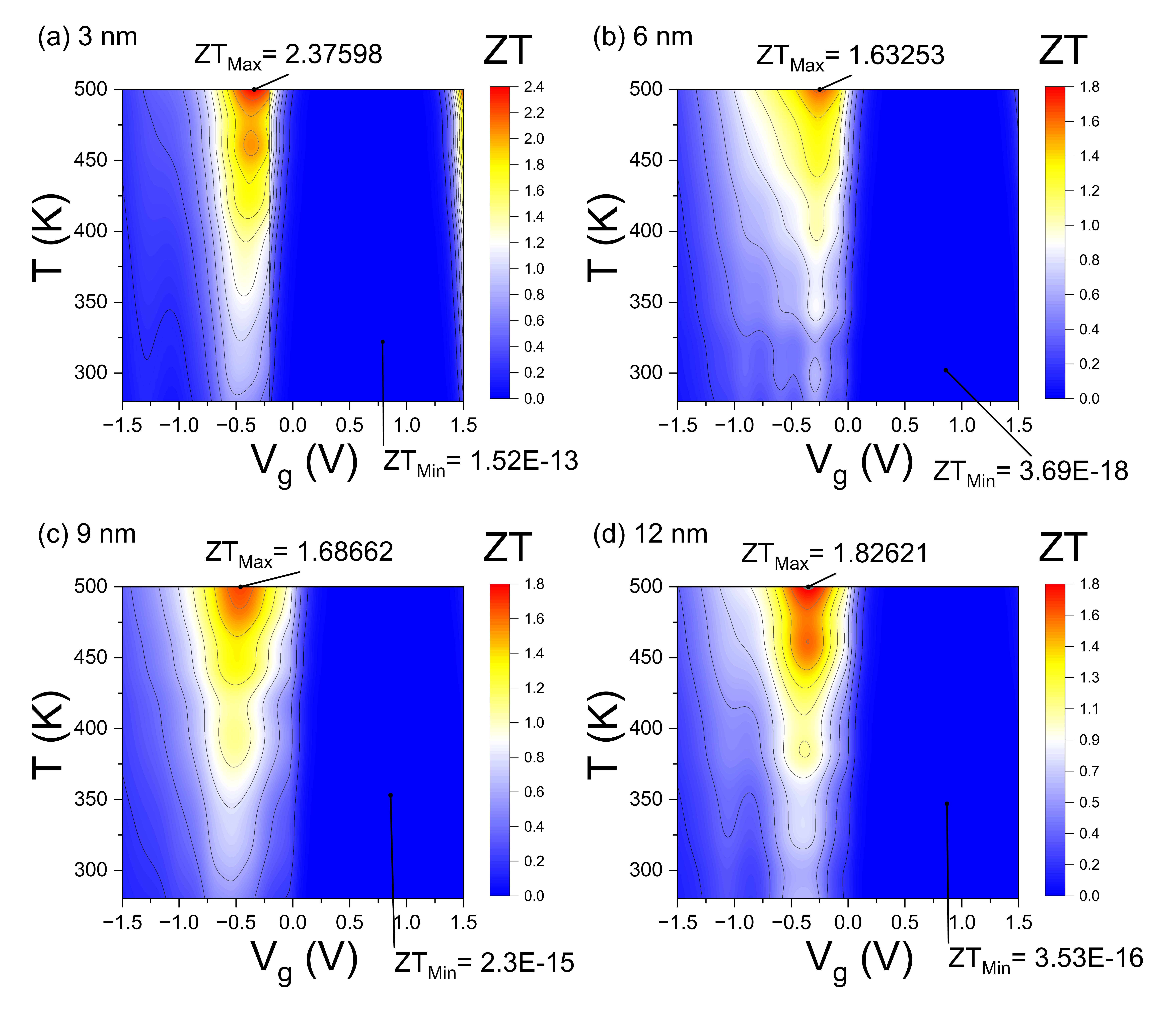}
\caption{ 
Contour plots of $ZT(T,V_g)$.
Contour plots of the thermoelectric figure of merit $ZT(T, V_g)$ are shown over the temperature range of 250–500 K and gate voltage range of –1.5 to 1.5 V for Pt–WSe$_2$–Pt thermoelectric junctions with channel lengths $\mathrm{L}_{\text{ch}}=$ (a) 3 nm, (b) 6 nm, (c) 9 nm, and (d) 12 nm. The maximum and minimum values of $ZT$, denoted as $ZT_{\mathrm{Max}}$ and $ZT_{\mathrm{Min}}$, respectively, along with their corresponding locations within the $T$–$V_g$ domain, are also indicated.
}
\label{fig:Fig8}
\end{figure*}

Figure~\ref{fig:Fig8} shows that the values of the factor $(1 + \kappa_{\mathrm{ph}}/\kappa_{\mathrm{el}})$ are enormous in the insulating regime, where $\kappa_{\mathrm{ph}} \gg \kappa_{\mathrm{el}}$. Notably, the extrema of the Seebeck coefficient $S$, which also correspond to the maxima of $S^2/L$, occur within this insulating regime. However, due to the extremely large values of $(1 + \kappa_{\mathrm{ph}}/\kappa_{\mathrm{el}})$ in these regions, the thermoelectric figure of merit $ZT$ is significantly suppressed, despite the large $S$. As a result, tuning $V_g$ to optimize the Seebeck coefficient does not necessarily lead to the optimization of $ZT$ in Pt–WSe$_2$–Pt thermoelectric junctions, because the denominator in the $ZT$ expression dominates and offsets the gains from an enhanced $S$.

Finally, the contour plots of $ZT(T, V_g)$ over the temperature range of 250–500 K and gate voltage range of –1.5 to 1.5 V are shown in Fig.~\ref{fig:Fig8}. Taking the 12 nm junction as an example, the maximum value of $ZT$ at room temperature is approximately 0.61 at $V_g = -0.39$ V. The value of $ZT$ further escalates with temperature, exhibiting local maxima of 1.09 and 1.62 at $T=$ 384 and 461 K, respectively. The maximum value of $ZT$, denoted as $Z_{\mathrm{Max}}$, is $ZT_{\mathrm{Max}} \approx 1.82$ in the contour plot range. The applied gate voltage modulates the semiconducting junction into the crossover region between insulating and conducting states, achieving the optimal values of $ZT$. Be aware that the $ZT_{\mathrm{Max}} \approx 2.37$ values obtained using the 3 nm nanojunction are the most efficiently designed. The optimal condition of $ZT$ is achieved in the shortest junctions when the gate voltage $V_g$ tunes the nanojunction to the transition point between conducting and insulating states, as well as the crossover from quantum tunneling to classical thermionic emission.

\section{Conclusion}
\label{sec:Conclusion}

Using first-principles approaches, we employ VASP, NANODCAL, and LAMMPS to investigate the transport mechanisms and thermoelectric performance of Pt–WSe$_2$–Pt nanojunctions with channel lengths of 3 nm, 6 nm, 9 nm, and 12 nm. These junctions, featuring edge-contact geometries, behave as p-type semiconductors with band gaps of approximately 2.65 eV. The application of a gate voltage ($V_g$) provides an additional degree of tunability, enabling modulation of the nanojunctions from insulating to conducting states.

Variations in channel length, temperature, and gate voltage give rise to a competition between two distinct electron transport mechanisms: quantum tunneling and semiclassical thermionic emission. At room temperature ($T = 300$ K), the 3 nm junction exhibits dominant quantum tunneling behavior in the insulating regime, where the gate voltage $V_g$ positions the chemical potential $\mu(V_g)$ within the band gap. As the temperature increases, the range over which quantum tunneling dominates becomes narrower. This quantum transport region is further suppressed with increasing channel length: at 300 K, the 9 nm junction shows quantum transport confined to a narrow mid-gap region, whereas in the 12 nm junction, the transport mechanism transitions fully to thermionic emission over the entire temperature range of 300–500 K. Consequently, the Seebeck coefficient image profiles $S(T, V_g)$ display qualitatively distinct behaviors for shorter channels ($L_{\mathrm{ch}} = 3$ and 6 nm) compared to longer ones ($L_{\mathrm{ch}} = 9$ and 12 nm), reflecting the underlying transition in dominant transport mechanisms.

Although $ZT$ scales with the square of the Seebeck coefficient ($S^2$), maximizing $S$ alone does not ensure optimal thermoelectric performance. For example, $|S|$ can exceed 5000~$\mu$V/K in the insulating state due to an exceptionally small transmission coefficient $\tau(E)$. However, thermoelectric efficiency remains low in this regime because of the high internal resistance associated with suppressed electrical conductivity. The $ZT$ value reaches its peak when $V_g$ shifts the chemical potential slightly outside the band gap, where the slope of $\tau(E)$ is appreciable. In this metal–insulator transition region, where the phononic thermal conductivity $\kappa_{\mathrm{ph}}$ becomes comparable to the electronic thermal conductivity $\kappa_{\mathrm{el}}$, $ZT$ can exceed 0.6 at room temperature and rise above 1.6 at $T = 500$ K. Notably, the 3 nm junction attains a $ZT$ of 2.37 at 500 K within this crossover regime. At this elevated temperature, the dominant transport mechanism transitions from quantum tunneling to classical thermionic emission.



\begin{acknowledgments}
The authors thank MOE ATU, NCHC,  National Center for Theoretical Sciences(South), and NSTC (Taiwan) for support under Grant NSTC 111-2112-M-A49-032-. also supported by NSTC T-Star Center Project: Future Semiconductor Technology Research Center under NSTC 114-2634-F-A49-001-. This work was financially supported under Grant No. NSTC-113-2112-M-A49-037-, and supported by NSTC T-Star Center Project: Future Semiconductor Technology Research Center under NSTC 114-2634-F-A49-001-, and also supported in part by the Ministry of Science and Technology, Taiwan. We thank to National Center for High-performance Computing (NCHC) for providing computational and storage resources.
\end{acknowledgments}

\bibliography{apssamp}
\end{document}